\documentclass[journal]{IEEEtran}
\usepackage{graphicx}
\usepackage{amsmath}
\usepackage{subfigure}
\usepackage{multirow}
\usepackage{array}
\usepackage{algorithm}
\usepackage{algorithmic}
\ifCLASSINFOpdf

\else
\begin{document}
\title{Evaluating Wireless Proactive Routing Protocols under Scalability and Traffic Constraints}
\author{\IEEEauthorblockN{N. Javaid$^{\$,\pounds}$, A. Bibi$^{\$}$, Z. A. Khan$^{\dag}$, U. Khan$^{\$}$, K. Djouani$^{\pounds,\ddag}$}\\ \vspace{0.4cm}
%    \IEEEauthorblockA{ \{nadeem.javaid,djouani@univ-paris12.fr\}}\\
        $^{\$}$Department of Electrical Engineering, COMSATS, Islamabad, Pakistan. \\
        $^{\dag}$Faculty of Engineering, Dalhousie University, Halifax, Canada.\\
        $^{\pounds}$LISSI, University of Paris-Est Creteil (UPEC), France.\\
        $^{\ddag}$F'SATIE, TUT, Pretoria, South Africa.

     }
\vspace{-2cm}

%
%% make the title area

\maketitle
\IEEEcompsoctitleabstractindextext{%
\begin{abstract}
%\boldmath
In this paper, we evaluate and analyze the impact of different network loads and varying no. of nodes on distance vector and link state routing algorithms. We select three well known proactive protocols; Destination Sequenced Distance Vector (DSDV) operates on distance vector routing, while Fish-eye State Routing (FSR) and Optimized Link State Routing (OLSR) protocols are based on link state routing. Further, we evaluate and compare the effects on the performance of protocols by changing the routing strategies of routing algorithms. We also enhance selected protocols to achieve high performance. We take throughput, End-to-End Delay (E2ED) and Normalized Routing Load (NRL) as performance metrics for evaluation and comparison of chosen protocols both with default and enhanced versions. Based upon extensive simulations in NS-2, we compare and discuss performance trade-offs of the protocols, i.e., how a protocol achieves high packet delivery by paying some cost in the form of increased E2ED and/or routing overhead. FSR due to scope routing technique performs well in high data rates, while, OLSR is more scalable in denser networks due to limited retransmissions through Multi-Point Relays (MPRs).
\end{abstract}

}

\maketitle
\IEEEdisplaynotcompsoctitleabstractindextext
\IEEEpeerreviewmaketitle
%\vspace{-0.5cm}
%section 1
\section{Introduction}
Routing is an essential but demanding objective in Wireless Multi-hop Networks (WMhNs). Routing protocols are responsible to calculate and tackle route (re)establishment during routing process. These protocols are divided into two main categories; reactive and proactive based upon their routing behavior. Reactive protocols start route calculation when request for data is arrived. While in proactive routing protocols, all nodes periodically keep attempting to be aware of their neighbors as well as of whole network topology.

Proactive protocols periodically compute information about links and routes. Thus, in this way delay is reduced. Whereas, these computations result in generation of higher routing load. In case of high node densities and high traffic rates, smaller bandwidth may cause drop rates. To optimize routing overhead for achieving less drop rates, proactive protocols implement some optimization mechanisms, which reduce routing overhead of pre-computations. To examine these techniques in proactive protocols, we select, Destination-Sequence Distance Vector (DSDV) [1], Fish-eye State Routing (FSR) [2] and Optimized Link State Routing (OLSR) [3]. Instead of giving details, we have given some features of the chosen proactive protocols in Table. 1. We further enhance efficiency of these mechanisms by modifying their original versions and analyze the performance of these protocols. For this analysis, we consider different scalabilities from $10$ to $100$ nodes and varying network loads; $2$, $4$, $8$, $16$ and $32$ \textit{packs/s}.

\begin{table}[ht]
\begin{center}
\tiny
\caption{Proactive Routing Protocol in Brief}
  \begin{tabular}{| m{0.5cm} | m{1.4cm} | m{1.3cm} | m{0.8cm} | m{1.4cm} | m{0.8cm}|}
  \hline
  %\multicolumn{6}{c}{Table.1. Proactive Routing Protocol in Brief} \\
%    \hline
    \textbf{Protocol} & \textbf{Distinct Feature} & \textbf{Calculation of Path} &\textbf{Forwarding of Packets}& \textbf{Flooding Control Mechanism}& \textbf{Overhead Reduction}\\ \hline
  \textbf{DSDV} &{Trigger and Periodic Updates}&{Distributed Bellman Ford (DBF) Algorithm}&{Hop-by-Hop Routing}&{Exchange the Topological Info. with Neighbors Only}&{Incremental Updates}\\ \hline
  \textbf{FSR} & {Multi-Scope Routing with Graded Frequency Mechanism}&{DBF Algorithm} &{Hop-by-Hop Routing}&{Graded Frequency mechanism}&{Fish-eye Technique} \\ \hline
  \textbf{OLSR} &{MPR}&{Dijkstra Algorithm}& {Hop-by-Hop Routing}&{Re-transmission of TC messages via MPRs}&{MPRs} \\ \hline

\end{tabular}
\end{center}
\end{table}
\normalsize
%\vspace{-1cm}
%Section 2
\section{Related Work and Motivation}
Layuan, L. \textit {et al.} [4], consider different perspectives of simulation models for MANETs. Furthermore, based on the performance parameters; delay, jitter, throughput, loss ratio, routing load and connectivity of Ad-hoc On-demand Distance Vector (AODV), DSDV, Dynamic Source Routing (DSR) and Temporally-Ordered Routing Algorithm (TORA) protocols are simulated in their work for $900s$ with variable scalability from $10$ to 100 nodes.

In [5], AODV and DSR are compared for Ad-hoc Networks using NS-2. Authors deduce that AODV and DSR better perform under high mobility situations than DSDV. DSR outperforms comparitive to AODV in less stress situations; smaller number of nodes and lower routing load and/or mobility.  However, they simulate AODV and DSR with only $10$ number of sources and low pause times, whereas, our study takes up to $40$ sources and with different traffic rates.

Behavior of three on-demand routing protocols; AODV, DSR and DYnamic MANET On-demand (DYMO), is compared in different network demands in MANETs in [6]. Authors select performance measuring metrics; throughput, packet delivery ratio and average end-to-end delay.

In [7], simulations are carried-out to evaluate the performance of three reactive protocols; AODV, DSR and DYMO and three proactive protocols; DSDV, FSR and OLSR.

In this paper, we enhance efficiency of the selected protocols and evaluate three original versions with three enhanced ones in varying densities of nodes and increasing flows of traffic.
DSDV triggers route updates for every change in active routes, and also periodically disseminates these updates through flooding. To minimize routing overhead of flooding, we have changed route settling value from $6$ to $7$, while trigger update time from $15s$ to $30s$. In FSR, two periodic updates are used to calculate routes. The periodic intervals for these scopes are too high to compute recent topological information. Moreover, routes are only updated through periodic updates due to absence of trigger updates.  Therefor,  to achieve frequent updates, we modify inner-scope interval value from $5s$ to $1s$, and outer-scope interval from $20s$ to $5s$. OLSR uses HELLO messages on routing layer to get information about links. Trigger updates are generated in case of expiration of $HELLO\_LOSS$ value. This value is too much high as compared to link layer feed-back mechanism, and is not suitable to provide quick convergence. Therefore, for convergence purpose, we enhanced HELLO and TC intervals from $2s$ and $5s$ to $1s$ and $3s$, respectively.

%section 3
\section{Routing Operations in DSDV, FSR, and OLSR}
These protocols necessarily implement some operations to maintain end-to-end paths. Subsections discuss the protocols with their maintenance operations.

\textit{A. Maintenance operations}

\textit{i. Monitoring of Link Status} operation is used to maintain recent information about link status with their neighbors in the network. If a node does not receive any link state message from a neighbor for a certain number of successive link state intervals, the link is assumed to be broken.

\textit{ii. Triggered Route Updates} are generated for every change in the link status to update the routing information across the network.

\textit{iii. Periodic Route Updates} are used by proactive protocols to calculate routes periodically. Unlike trigger route updates, periodic route updates accumulate all information regarding link changes after a specific period of time.

\textit{B. Route Maintenance Operations in DSDV, FSR and OLSR}
All of the aforementioned three operations are performed in DSDV. Although, trigger route updates may appear redundant because of employment of periodic maintenance of links via link state updates. Updating the status of links with trigger updates may lead to routing loops in DSDV. So, periodic route updates with transmission of destination sequence numbers monitor and maintain freshness of the routes. A moderate approach is used in FSR, where trigger
updates are not performed at all. Drawback of using both link state monitoring and trigger updates cause large amount of control traffic generation. As, trigger updates are exchanged on every change in link status, they generate large number of routing messages, especially during the high rates of mobility. One of the challenges of using periodic route updates (with periodic link state monitoring) is to address the trade-off between amount of control traffic and the consistency in route information. OLSR performs only trigger updates to maintain fresh routes by using Topology Control (TC) messages along with link status monitoring through HELLO messages.

A protocol $(pro)$ has to pay some cost $C$ in the form of consumed energy $C_E$ and routing delay $C_T$ [8].

%eq1
\small
\begin{equation}
C_{pro}(\mu)=C_E(\mu)\times C_T(\mu)
\end{equation}
\normalsize

Where, $\mu$ is task oriented input data, i.e., either number of nodes, or number of broken links, or number of sent packets, or limit of number of packets in buffer queue, etc. Different costs (price to pay) make suitable a protocol for different situations. We use the term $C_p$ alternative to $C_E$ for measuring energy cost in terms of routing packets.

Total routing overhead cost $C_p^{DSDV}$ presented in eq.2 and eq.3 depend on $C_p^{per}$ and $C_p^{trig}$. $\tau_{per}$ is periodic exchange interval, $\tau_{NL}$ is total network life time and $Change_i^{ActiveRoute}$ shows change in link, $i$, among active route.

%eq2
\small
\begin{equation}
C_p^{DSDV}=C_p^{per}+C_p^{trig}
\end{equation}
\normalsize

%eq3
\small
\begin{equation}
C_p^{DSDV}=\frac{\tau_{NL}}{\tau_{per}}\sum_{\forall i\in N}i+\int^{\tau_{NL}}|Sgn(Change_i^{AtiveRoute})|\sum_{\forall n\in N}n
\end{equation}
\normalsize

If $i$ in an $ActiveRoute$ changes then $|Sgn(Change_i^{AtiveRoute})=1$, otherwise it is $0$.

Eq.4 and eq.5 describe total routing cost of FSR; $C_p^{FSR}$, which is sum of the packet cost for dissemination in inner-scope, $ C_p^{per-in}$, and in outer-scope $ C_p^{per-out}$.

%eq4
\small
\begin{equation}
C_p^{FSR}=C_p^{per-in}+C_p^{per-out}
\end{equation}
\normalsize

%eq5
\small
\begin{equation}
C_p^{FSR}=\tau_{NL}\left(\frac{1}{\tau_{in}}\sum_{\forall i\in N_{in}}i+\frac{1}{\tau_{out}}\sum_{\forall j\in N_{out}}j\right)
\end{equation}
\normalsize

In eq.5, $ \tau_{in}$ and $\tau_{out}$ parameters are used for inner-scope and outer-scope intervals for periodic route updates, while, $N_{in} $ and $ N_{out} $ are the nodes in inner-scope and outer-scope.

The packet cost of OLSR; $C_p^{olsr}$ is measured in eq.6, which is the sum of periodic HELLO messages' cost; $C_p^{hello}$, trigger cost of TC messages due to MPR redundancy $C_p^{TC-trig}$ and default cost of TC messages due to the stable MPRs $C_p^{TC-def}$. Whereas, $\tau_{hello}$ specifies HELLO interval. Moreover, we have defined three sets of nodes; \textit{(i)} connected neighbor nodes; $N_{br}$, \textit{(ii)} selected MPRs, $MPRs$, and \textit{(iii)} all nodes in the network $N$.

\small
\begin{equation}
C_p^{OLSR}=C^{hello}_p+C^{TC-trig}_p+C^{TC-def}_p
\end{equation}
\normalsize

where,

\small
\begin{equation}
C^{hello}_p=\frac{\tau_{NL}}{\tau_{hello}}\sum _{ \forall i \in N} \sum_{\forall j \in Nbr}j
\end{equation}
\normalsize

\small
\begin{equation}
C_p^{TC-trig}=\int^{\tau_{NL}}\sum_{\forall i\in N}\sum_{\forall j\in MPRs}|Sgn(Change_j^{MPR})|j
\end{equation}
\normalsize

\small
\begin{equation}
C_p^{TC-def}=\int^{\tau_{NL}}\sum_{\forall i\in N}\sum_{\forall j\in Nbr}|Sgn(Change_j^{MPR})|j
\end{equation}
\normalsize

The trigger updates of OLSR depend upon $Change_j^{MPR}$.

\section{Modeling Routing Overhead of Proactive Protocols}
For DSDV, expected energy cost to be paid in the form of number of generated routing packets in [9] is:

\small
\begin{equation}
U^{DSDV}\emph{}=N\times \tau_{NL}\times \alpha
\end{equation}
\normalsize

Where, $U$ is utilization metric, $N$ is number of nodes, and $\tau_{NL}$ is network life time and $\alpha$ is rate of route table advertisement including trigger and periodic updates, as shown in Fig.1.

\begin{figure}[h]
  \centering
   \subfigure{\includegraphics[height=7.3 cm,width=6 cm]{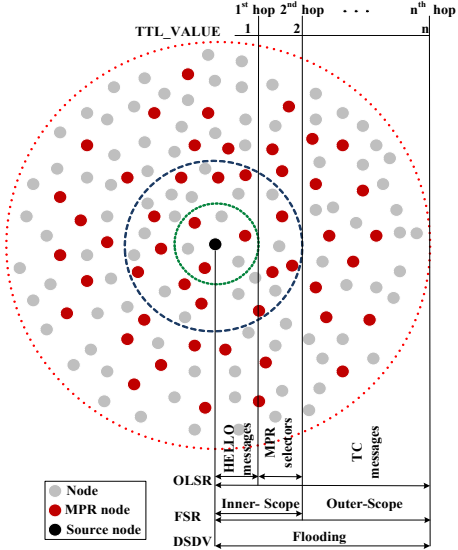}}
  \caption{Flooding, Source Routing and MPR Routing}
\end{figure}

In FSR, instead of flooding (as in DSDV), route updates are broadcasted in InterScope and IntraScope. $\alpha_{per}^{in}$ and $\alpha_{per}^{out}$ are rates of periodic updates in inner and outer-scopes, respectively. $N_{in}$ denotes number of nodes in inner-scope and $N_{out}$ are number of nodes in outer scope.

\small
\begin{equation}
U^{FSR}=U^{in}_{per}+U^{out}_{per}
\end{equation}
\normalsize

\small
\begin{equation}
U^{FSR}=N_{in}\times \tau_{NL}\times \alpha^{in}_{per}+N_{out}\times\tau_{NL}\times\alpha^{out}_{per}
\end{equation}
\normalsize

In OLSR, flooding takes place through MPRs. Each two-hop path is evaluated in terms of $U$ for MPR mechanism [10], $U_{\eta_2}^{MPR}$. It is calculated for a path from source node, $\eta_s$ towards its two-hop neighbors $\eta_2$ through a relay node $\eta_1$ as follows:

\small
\begin{equation}
U_{\eta_2}^{OLSR-MPR}=\frac{B_f\times EU}{D}
\end{equation}
\normalsize

Where, $B_f=\frac{B_A}{B_a}$ is a bandwidth factor between nodes $\eta_s$ and $ \eta_1(MPR),B_A$  is a available (free) bandwidth at $ \eta_1$, $B_a$ is an expected/requested outgoing bandwidth at the source node $\eta_s$. $EU=\frac{E_A^{\eta_1}}{E_{Tx}^{\eta_1\to\eta_2}}$ is the cost metric between
$\eta_s$ and its two-hop neighbor $ \eta_2$, $ E_A^{\eta_1}$ is an available energy at $ \eta_1$ in joules, $ E_{T_x}^{{\eta_1} \to {\eta_2}}$ is an energy used to transmit messages from $ \eta_1$ to $ \eta_2$, and $D$ is an end-to-end delay from $\eta_s$ to $ \eta_1$ in seconds.

In the next section, practical evaluation of selected protocols is discussed in detail.

\section{Simulation Setup}
In high traffic rates and densities, delay results in drop rates. Proactive protocols minimize routing delay due to pre-computation of routes. For assessment of these protocols, we select different traffic rates and scalabilities using NS-2. For scalability analysis, number of nodes are varied from $10$ to $100$ with packet size of $512 bytes$.  For different traffic rates, $2$, $4$, $8$, $16$, and $32 packs/s$ are selected for $50$ nodes, whereas, size of the packet is set to $64 bytes$. To examine the behavior of protocols for both selected scenarios, simulations are run for $900 s$ for packet with speed of $20m/s$ with pause time of $2s$. The sources transmit Continuous Bit Rate (CBR) traffic. Bandwidth provided to all the wireless links is $2\,\,Mbps$. The nodes taking part  simulation are randomly dispersed in an area of $1000m \times 1000m$ using Random Way-point Model.

\section{Evaluating Proactive Routing Protocols}
Performance of the protocols has been evaluated and compared with three performance parameters; throughput, E2ED, and NRL.

\begin{figure*}[!t]
  \centering
   \subfigure[Traffic: Original Protocols]{\includegraphics[height=3 cm,width=5.5 cm]{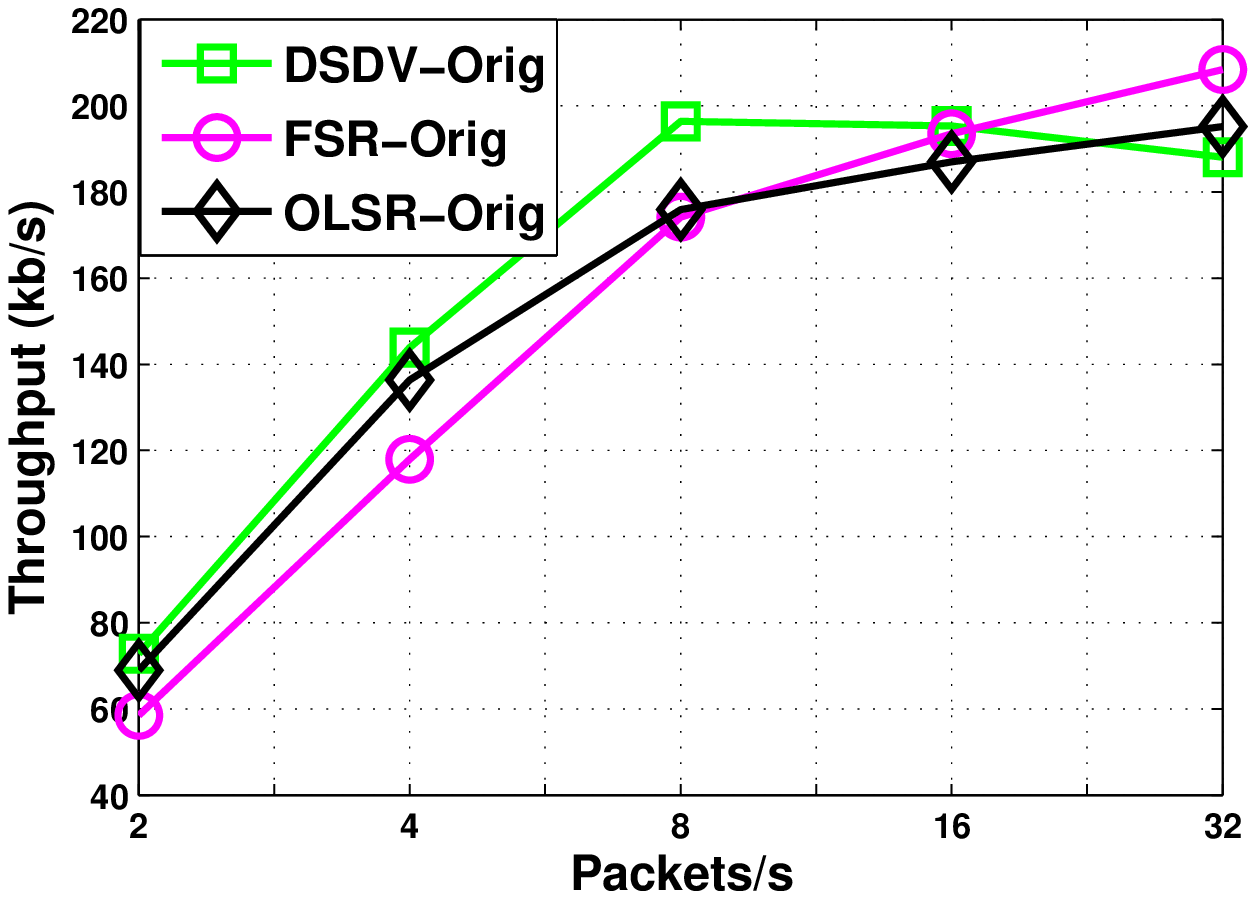}}
    \subfigure[Traffic: Original Protocols]{\includegraphics[height=3 cm,width=5.5 cm]{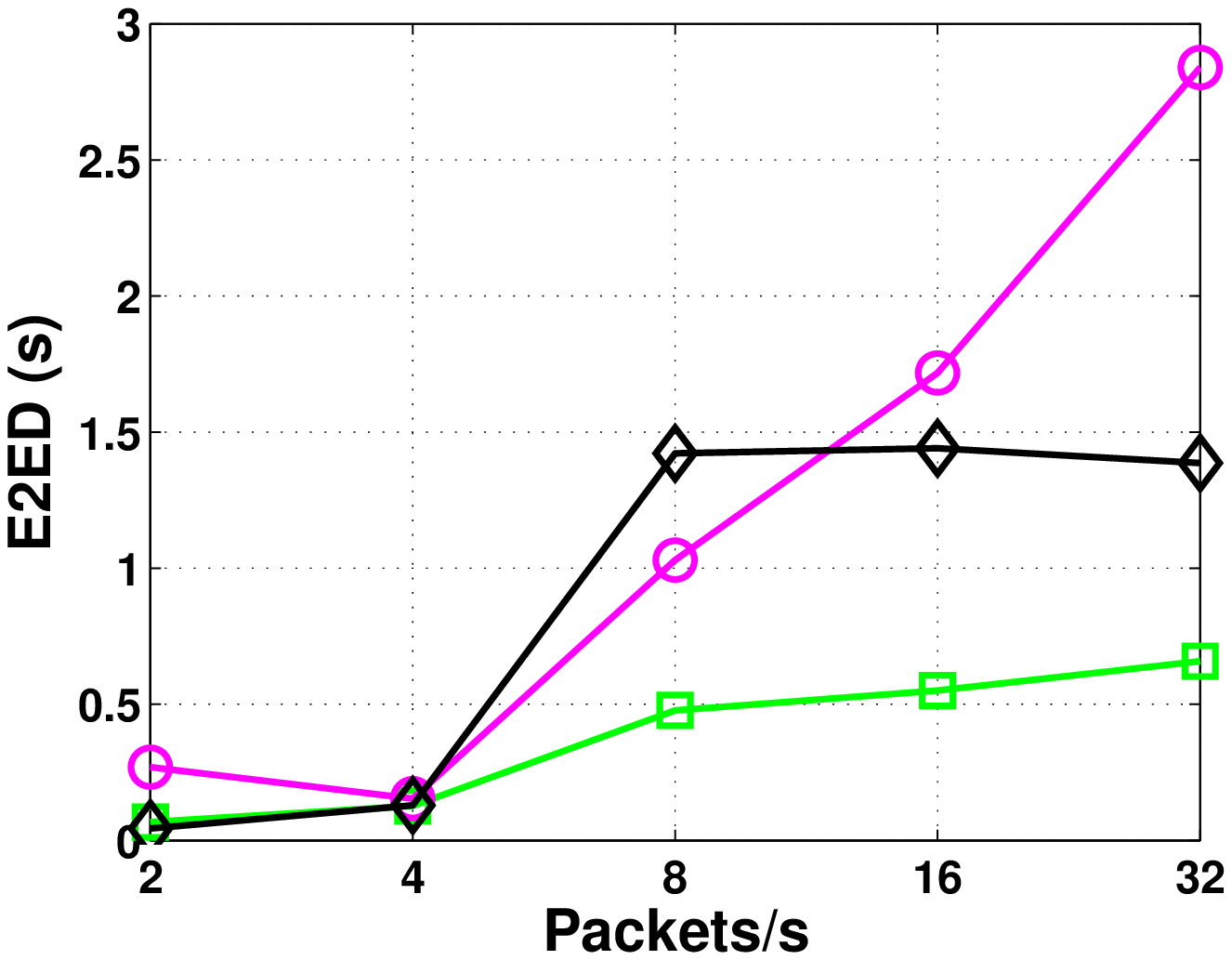}}
    \subfigure[Traffic: Original Protocols]{\includegraphics[height=3 cm,width=5.5 cm]{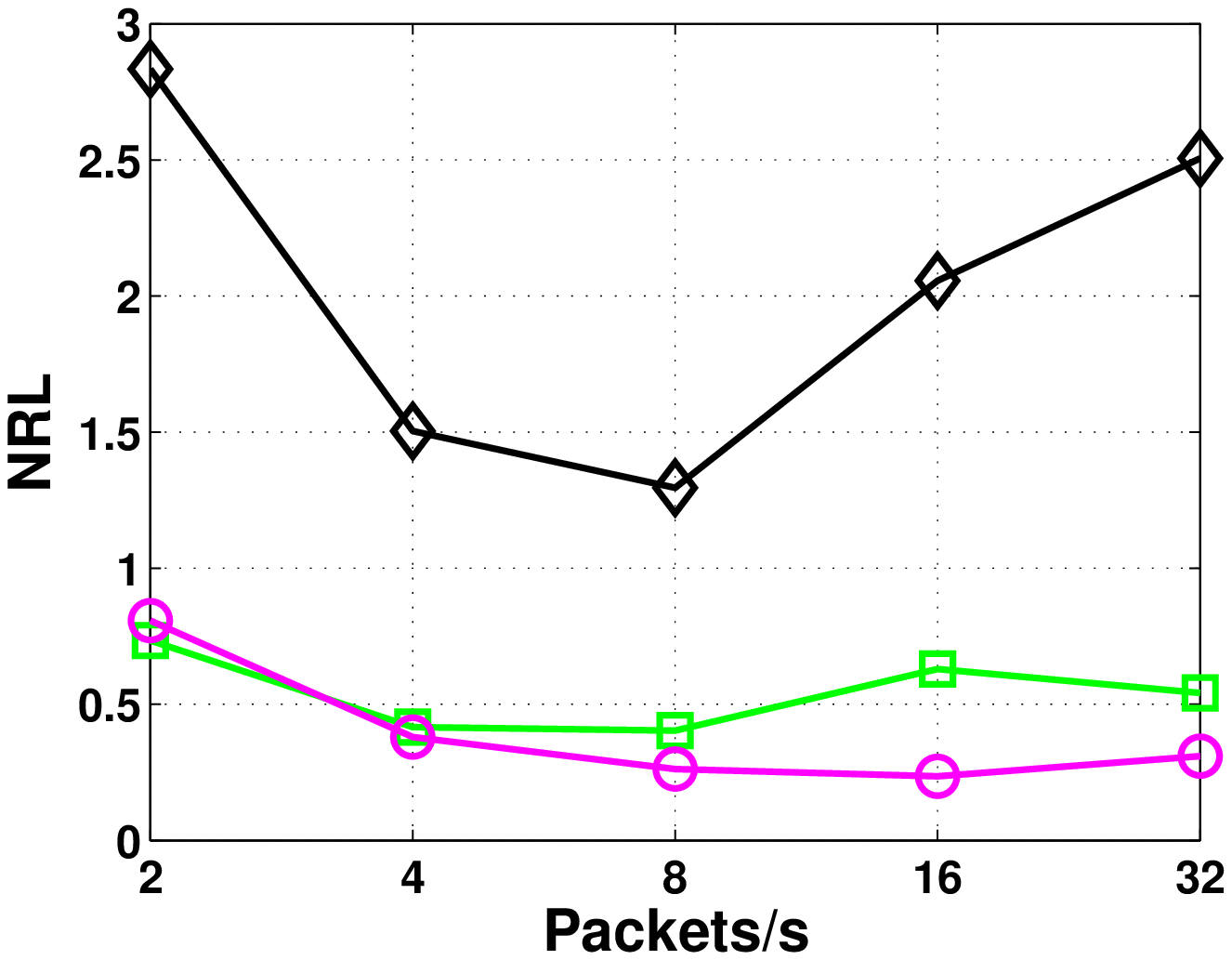}}

 \subfigure[Traffic: Modified Protocols]{\includegraphics[height=3  cm,width=5.5 cm]{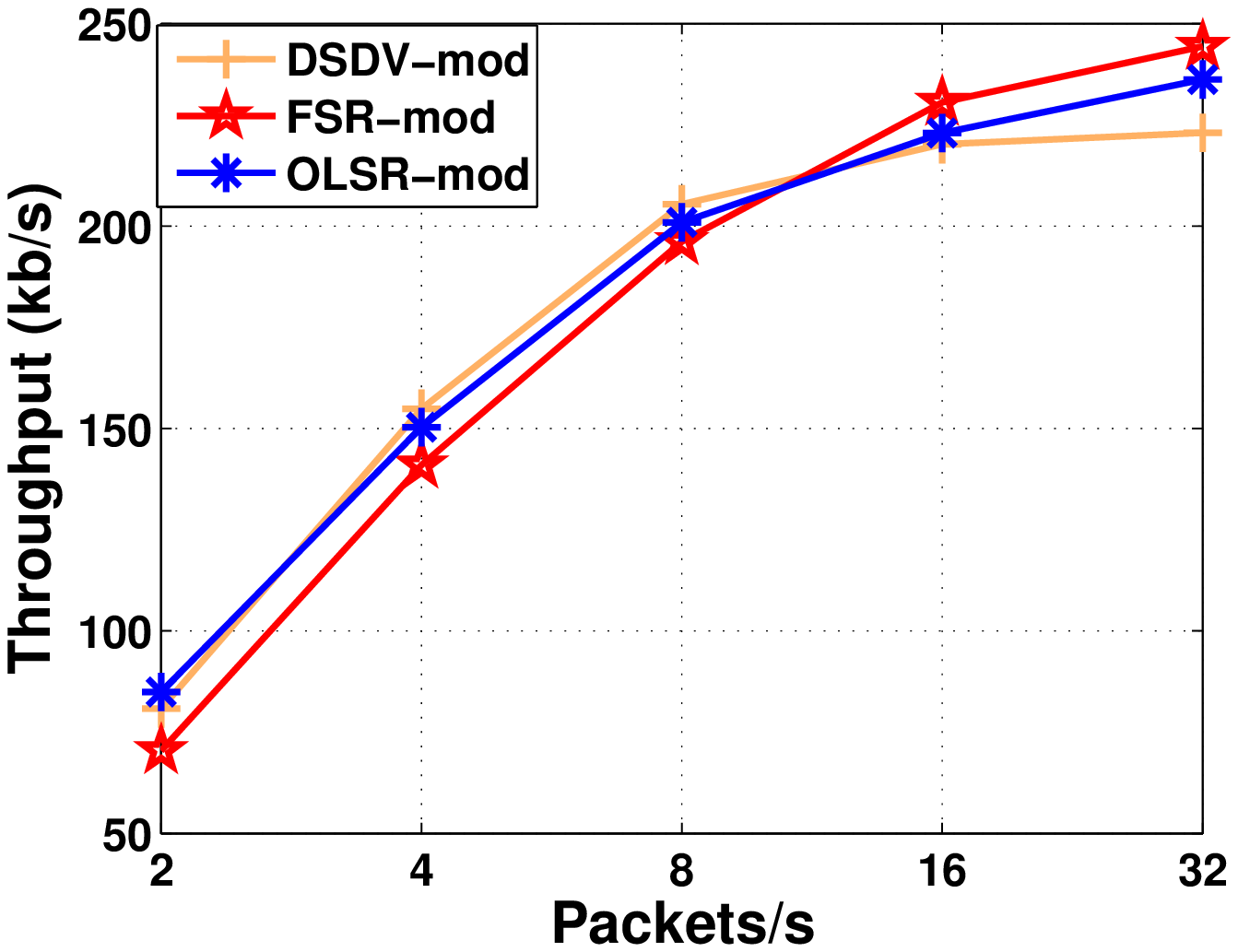}}
  \subfigure[Traffic: Modified Protocols]{\includegraphics[height=3  cm,width=5.5 cm]{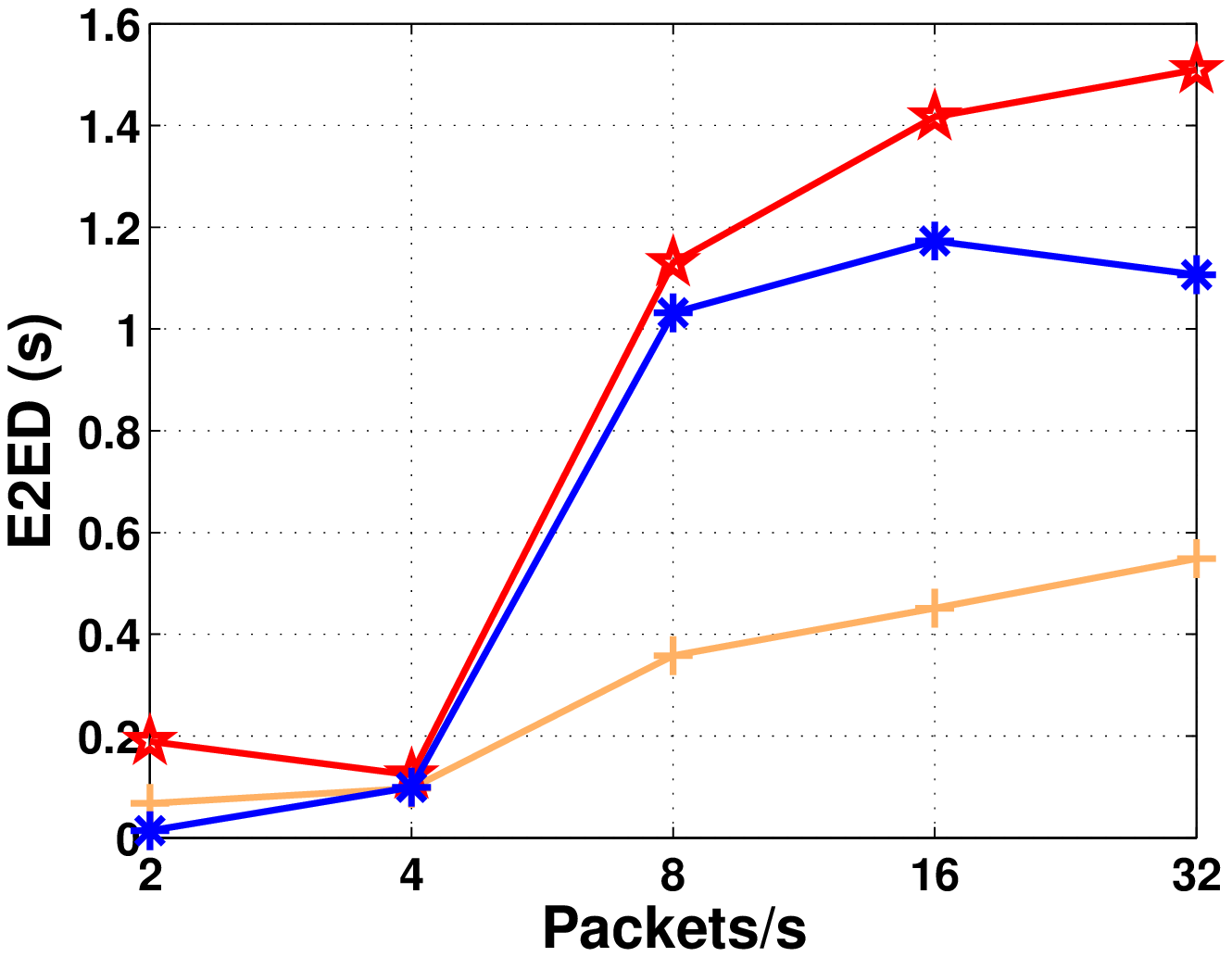}}
  \subfigure[Traffic: Modified Protocols]{\includegraphics[height=3  cm,width=5.5 cm]{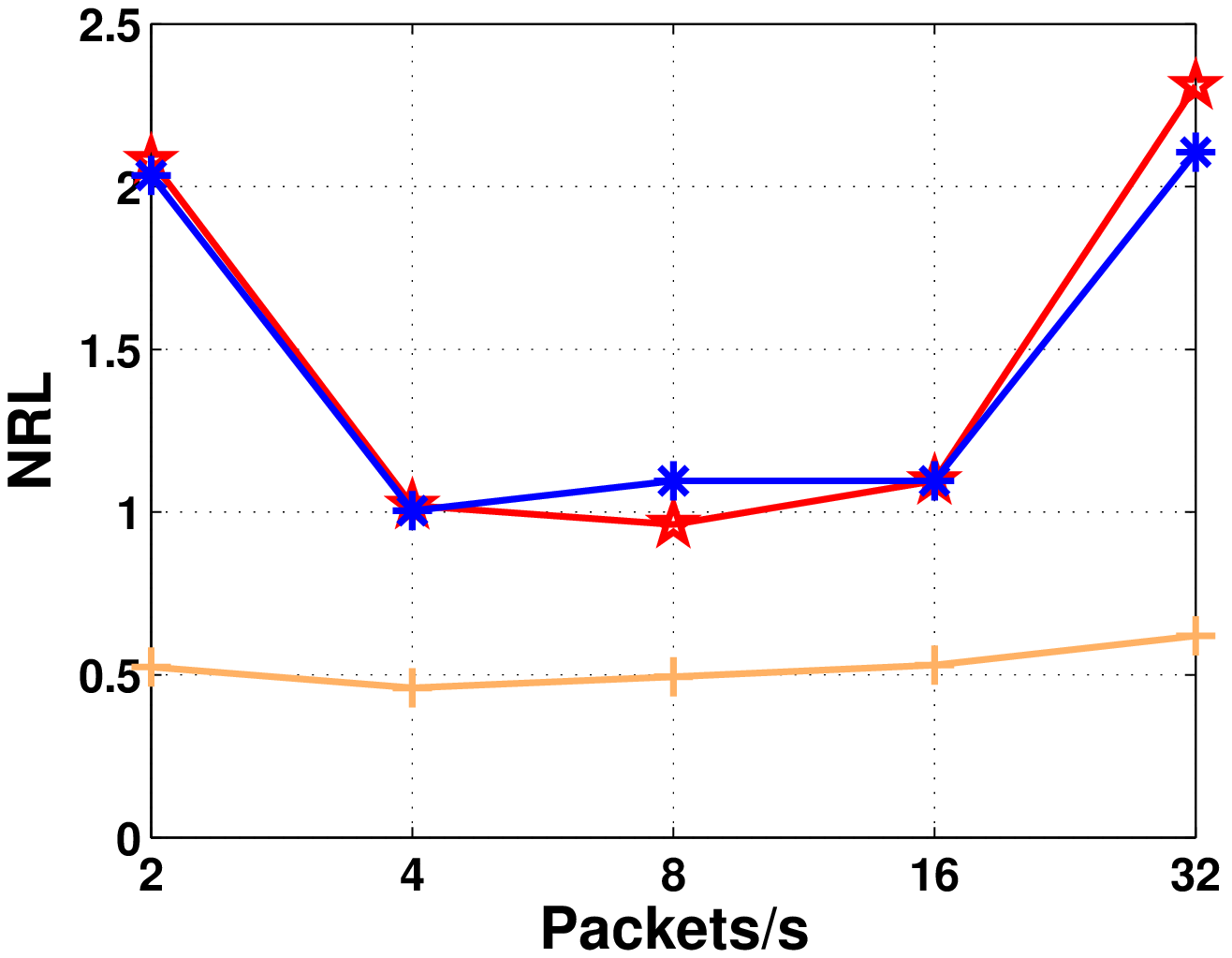}}

 \subfigure[Scalability: Original Protocols]{\includegraphics[height=3 cm,width=5.5 cm]{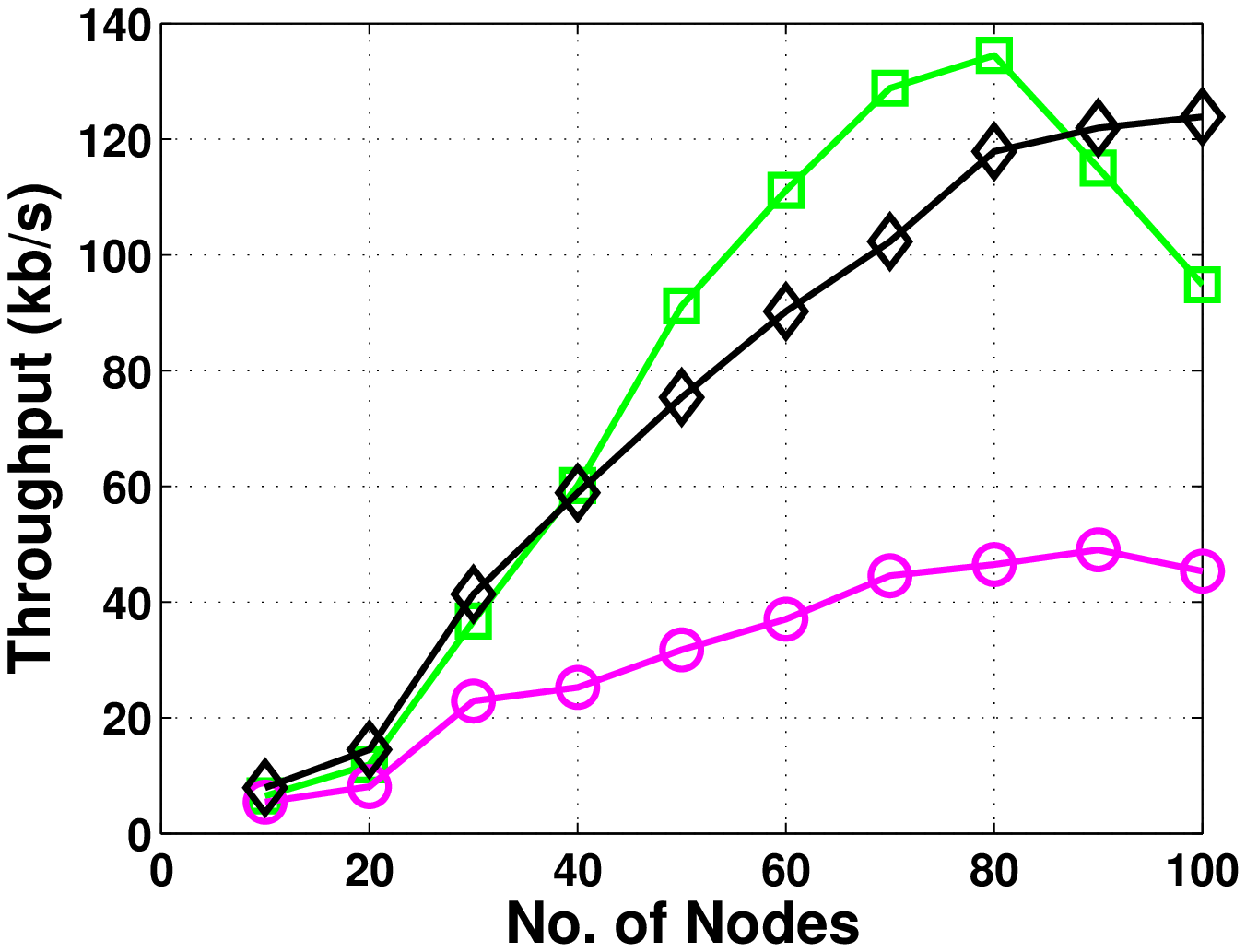}}
  \subfigure[Scalability: Original Protocols]{\includegraphics[height=3 cm,width=5.5 cm]{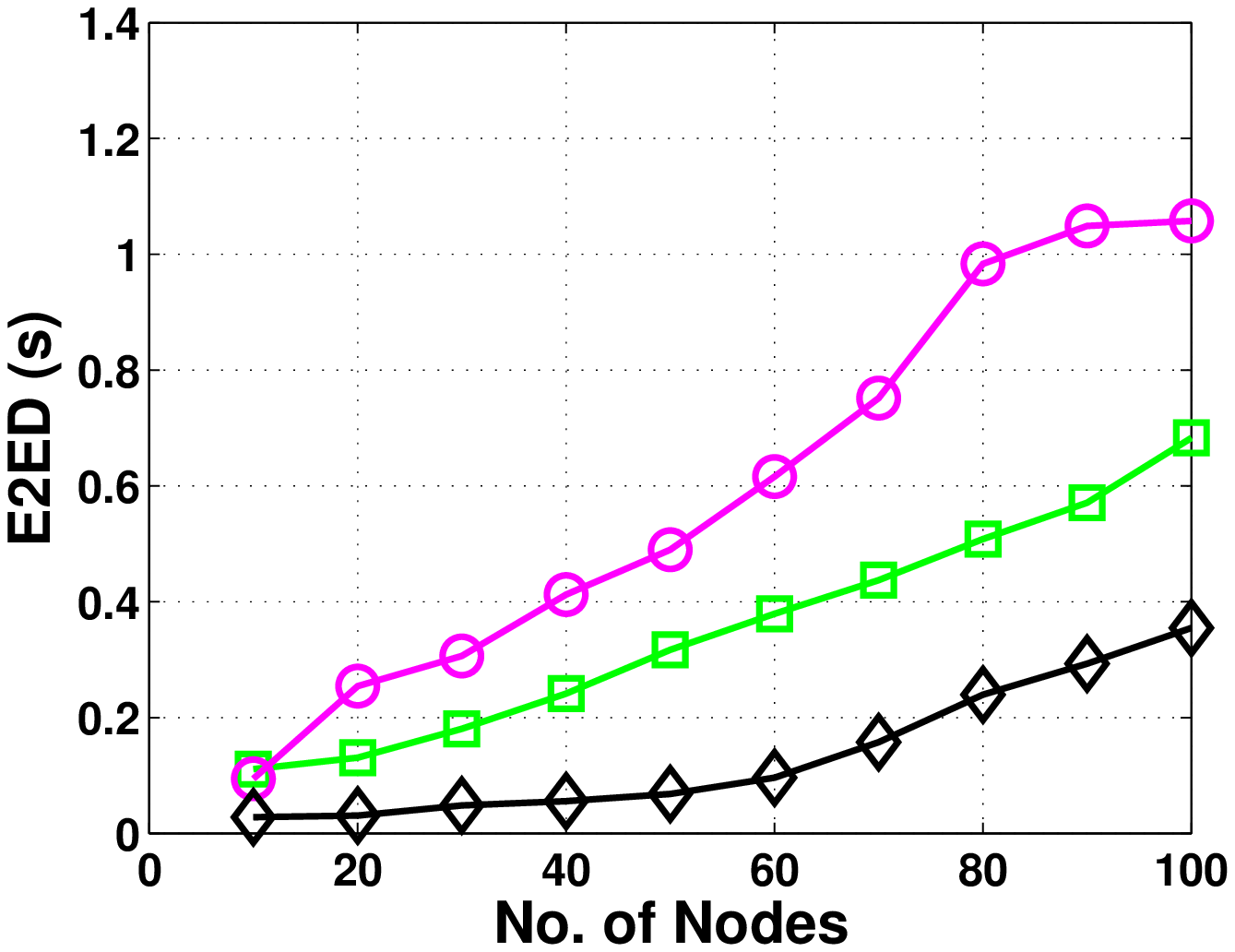}}
  \subfigure[Scalability: Original Protocols]{\includegraphics[height=3 cm,width=5.5 cm]{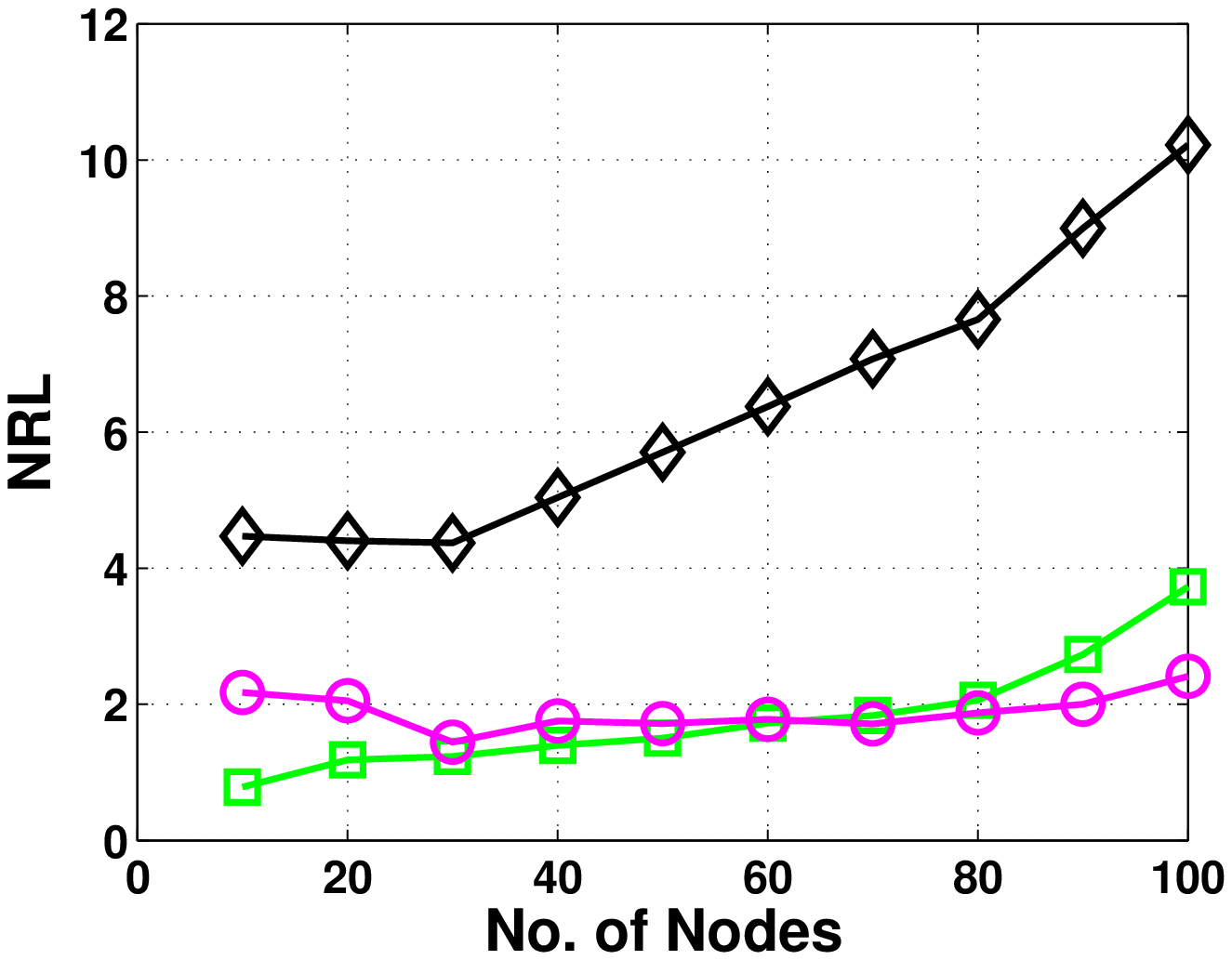}}

 \subfigure[Scalability: Modified Protocols]{\includegraphics[height=3  cm,width=5.5 cm]{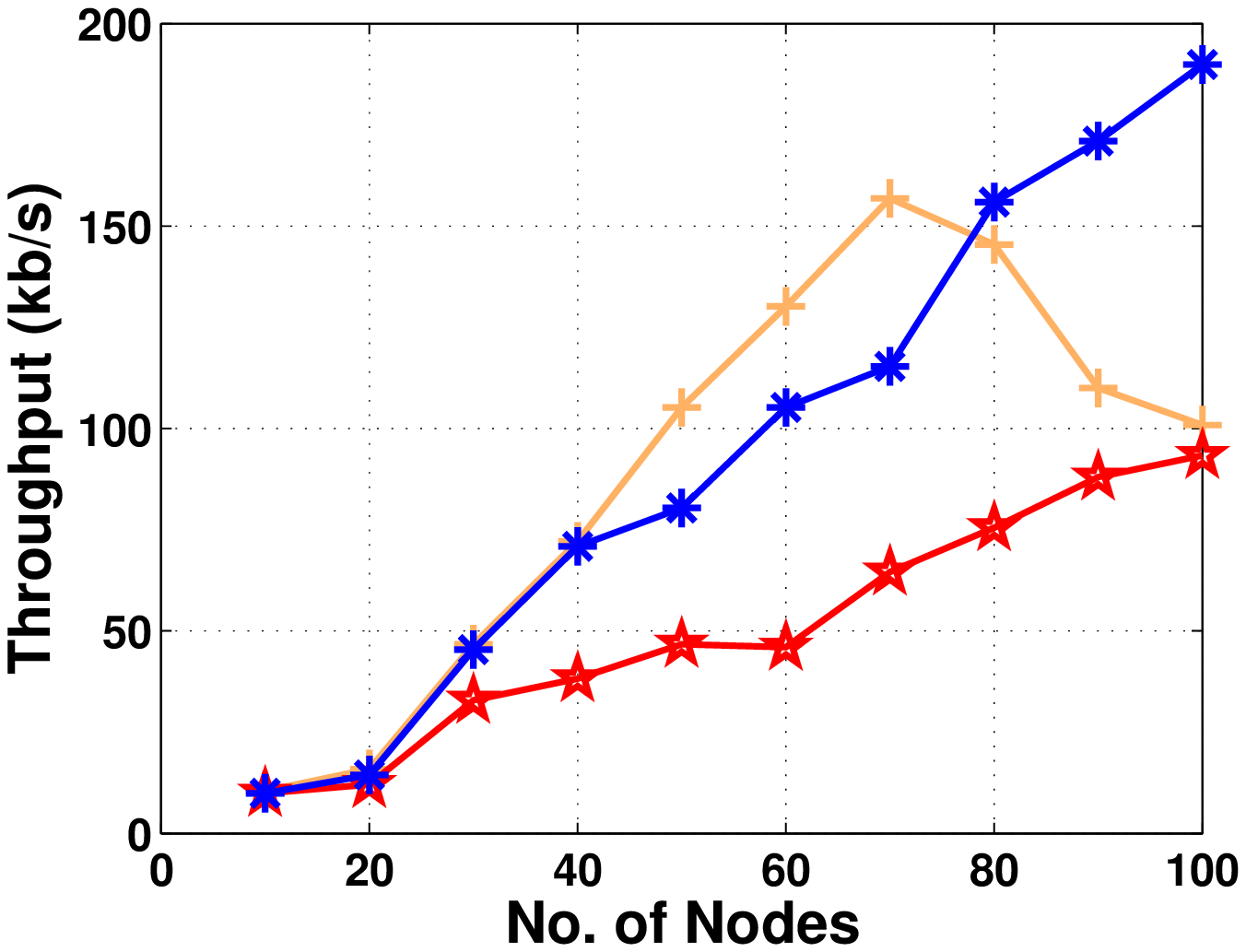}}
 \subfigure[Scalability: Modified Protocols]{\includegraphics[height=3  cm,width=5.5 cm]{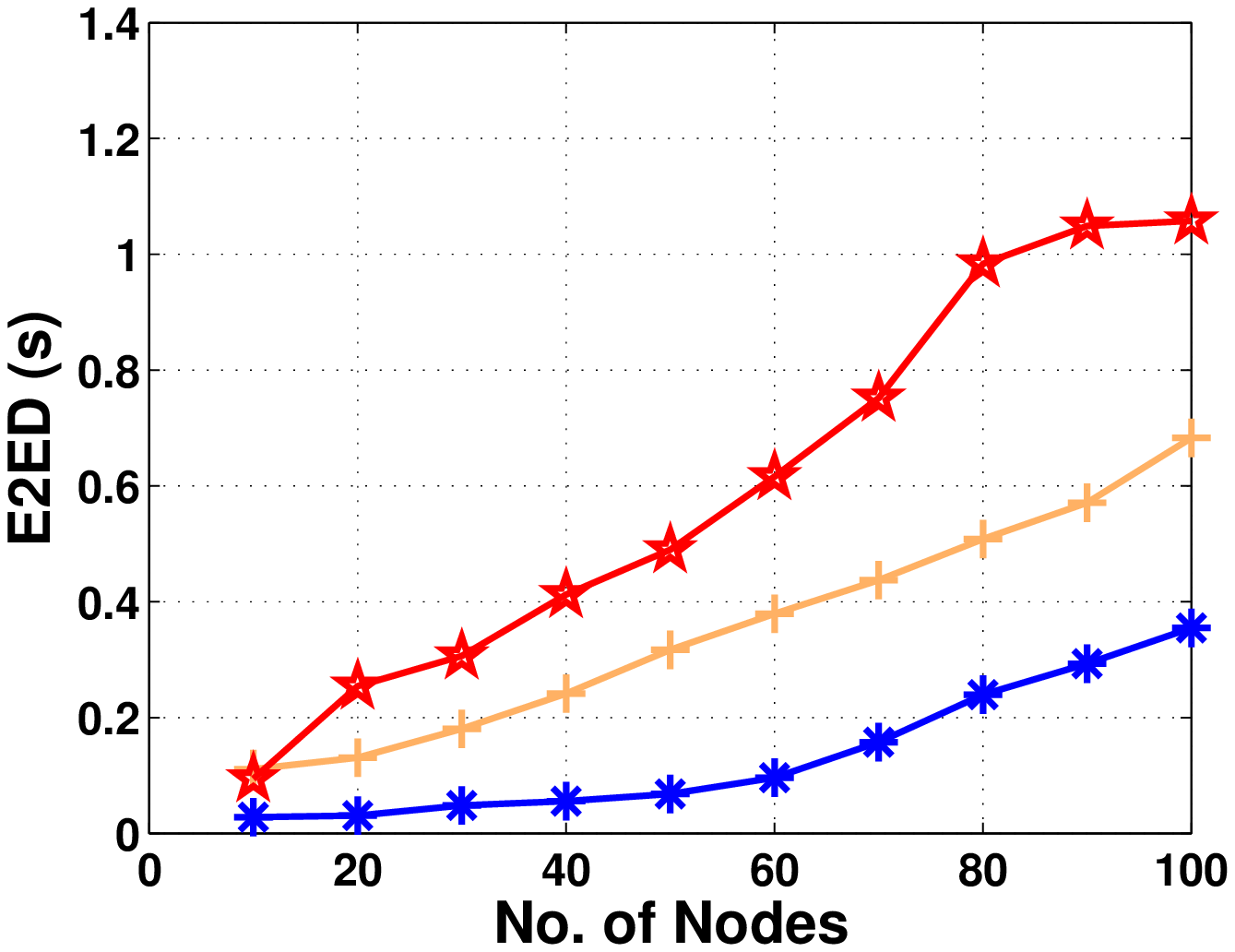}}
  \subfigure[Scalability: Modified Protocols]{\includegraphics[height=3  cm,width=5.5 cm]{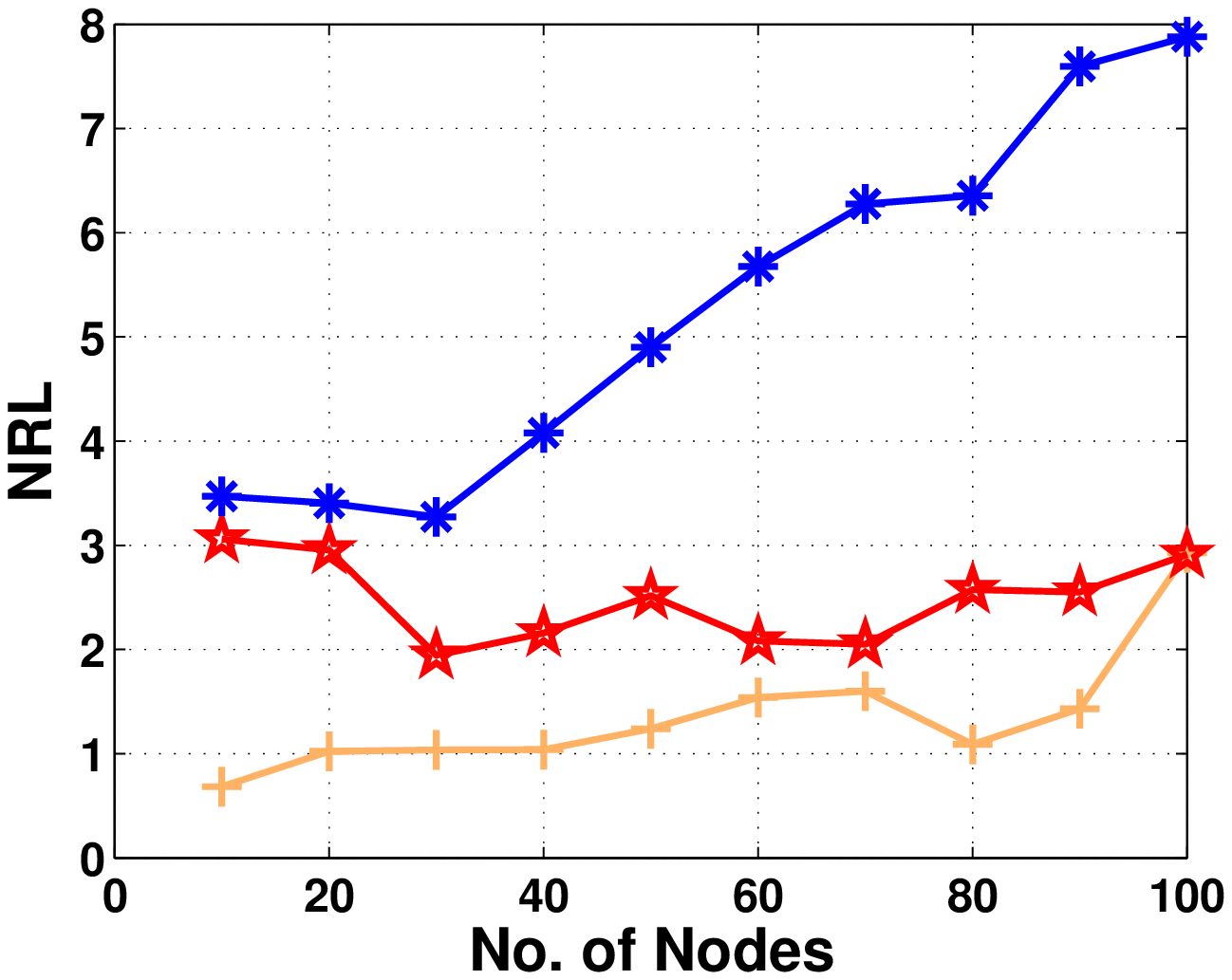}}

  \caption{Simulations Carried-out for this Study}
\end{figure*}

\textit{A. Throughput}
is amount of data successfully transferred from sources to destination during the specified simulation time. Our enhancements increase throughput of all selected protocols due to reduction in routing overhead. As, overall behavior of the chosen protocols remains same, thus, we use DSDV, FSR and OLSR as general term for presenting both original and enhanced versions. FSR shows appreciable performance for varying traffic rates and OLSR is well scalable among proactive protocols (Fig.2). In medium and high traffic loads, FSR's efficiency is depicted in Fig.2.a. This is due to introduction of new technique of multi-level Fish-eye Scope (FS), that reduces routing overhead and works better when available bandwidth is low, thus increasing throughput in case of increased data traffic loads and reduces routing update overhead. Although, DSDV uses Network Protocol Data Units (NPDUs) to reduce routing transparency but trigger updates cause routing overhead and degrade performance. OLSR uses MPRs for reduction of overhead but computation of these MPRs takes more bandwidth. Therefore, its throughput is less than FSR. Moreover, through updating link state information with different frequencies depending on FS distance, FSR well scales to large sized networks. FS technique allows for exchanging link state messages at different intervals in a network within different FS distances that reduce the link state message size. Further optimization helps FSR to only broadcast topology messages to neighbors in order to reduce flood overhead. If FSR would have taken MAC layer feedback in case of link brakes then there might be increased exchange of messages to update neighbors, consuming bandwidth and lowering throughput. This faster discovery results in a better performance during high traffic loads (Fig.2.a,b).

Simulation results of OLSR in Fig.2.a,b,c,d show that it is scalable but less converged protocol for high traffic rates. This protocol is well suited for large and dense mobile networks, as it selects optimal routes (in terms of number of hops) using MPRs. MPR computation is used to reduce dissemination overhead which produces typical flooding process, thus occupies precious bandwidth and drops the data packets. In a dense network, more optimizations can be achieved as compared to the classic link state algorithm. MPRs better achieve scalability in the distribution of topology information.

In higher network flows (scalabilities as well as mobilities), DSDV's throughput is decreased, as shown in Fig.2.a,b,c,d, which is increasing throughput ratio New route entry is advertised in DSDV when a subsequent forwarding data packet is requesting for new destination. This advertisement leads to increase in routing overhead, thus decreases throughput. In high scalabilities, DSDV produces lower throughput, as route settling time increases average E2ED and multiple NPDUs, increase routing load in large population. Moreover, NRL increases due to occurrence of more full dumps (changing the entire routing table) that consequently affects packet deliveries.

\textit{Interesting Facts Regarding Throughput:}
While considering throughput, routing load is more important issue to be tackled in high data traffic; freer bandwidth is demanded by the data requests. As, \textit{(1)} Proactive protocols  periodically compute routes to reduce routing latency but it augments routing load. \textit{(2)} In DSDV, trigger updates along with periodic updates cause more routing overhead. In high scalabilities, active routes are also increased. Any change in an active routes produces routing overhead due to trigger updates.

FSR reduces routing load via scope routing (no flooding) while updating the routes periodically. Thus in high data traffic rates, more bandwidth is available for data which increases throughput. In high scalabilities, network is more congested and demands for low latency. To reduce (re)transmissions caused by longer routes, instead of using simple flooding OLSR introduces MPR mechanism.

\textit{B. E2ED}
is the time a packet takes to reach the destination from the source.
Increase in traffic rates and node density result more delay for all of three proactive protocols. FSR overall suffers higher delay in both situations, (Fig.2.e,f,g,h). To retain route entries for each destination, this protocol maintains low single packet latency when traffic load or population is small. The graded frequency (GF) mechanism is used to find destination to keep routing overhead low. FSR exchanges updates more frequently to the near destinations. Thus, in higher data rates or more scalabilities this protocol attains more E2ED. The reason for delay in DSDV is that it waits to transmit a data packet for an interval between arrival of first route and the best route. Thus, this
selection introduces delay in advertising routes which are about to change. A node uses new entry for subsequent forwarding decisions and route settling time is used to decide how long to wait before advertising it. This strategy helps to compute accurate route but produces more delay. A proactive protocol first calculates routing tables, so, for larger networks, it takes more time resulting in more E2ED. Small values of AE2ED for OLSR are seen among proactive
protocols in all scalabilities, as shown in Fig.2.e,f,g,h, because, MPRs provides efficient flooding control mechanism, i.e., instead of broadcasting, control packets are exchanged with neighbors only.

\textit{Interesting Facts Regarding E2ED:}
In high scalabilities, (re)transmissions through relay nodes due to longer paths in a network need delay reduction. Flooding mechanism in DSDV along with route settling time introduce more delay, as processing of routing information of intermediate nodes increase latency as compared to OLSR which excellently reduces (re)transmission latency. Scope-routing is one of efficient algorithms to reduce routing overhead, but scope period for refreshing route updates of FSR have more interval between successive scopes updates; inner-scope period $=5s$ and outer-scope
period $=15s$. Thus its routing latency is much more than OLSR and DSDV in high data loads along with all scalabilities. OLSR achieves the lowest E2ED because of three distinguished features; \textit{(1)} It implements lowest update routing interval as compared to remaining two protocols; $2s$ in original and $1s$ for enhanced versions for link state monitoring through HELLO messages and $5s$ in original and $3s$ for enhanced versions for TC messages to update routes. Whereas, DSDV has interval of 15s for route updates, and in FSR, interval for inner-scope is of $5s$ and for outer-scope is $15s$, \textit{(2)} reduces (re)transmission latency through MPRs, and \textit{(3)} OLSR updates routes by trigger updates through TC redundancy in case of high dynamicity.

\textit{C. NRL}
is the number of routing packets transmitted by a routing protocol for a single data packet to be delivered successfully at destination.
As depicted in Fig.2.i,j,k,l, in all scalabilities and traffic loads, OLSR is generating the highest NRL. It happens due to MPR mechanism that controls the dissemination of control packets in the whole network. But calculation of these MPRs through TC and HELLO messages increase routing load. Moreover, OLSR link state messages are used to calculate MPRs that generate routing overhead. DSDV and FSR sustain low overhead in all network loads and in low and medium scalabilities. As, DSDV upholds routing table with separate route entry for new destination, while a node does not use the new entry for the same destination in making subsequent forwarding decisions. Moreover, NPDUs are arranged to disseminate incremental updates for maintaining low routing overhead.\newpage

\textit{Interesting Facts Regarding NRL:}
Routing load depends upon interval between routing updates; shorter the interval more routing load. As, OLSR generates routing updates in shorter interval as compared to rest of the protocols, it produces the highest NRL in both OLSR-Orig and OLSR-Mod. Trigger updates generate more routing load as compared to periodic routing updates. Both DSDV and OLSR generate trigger updates, but DSDV triggers routing updates only for link breakage among active routes as compared to OLSR that generate trigger updates for every change in the links. FSR uses only periodic updates, moreover, scope routing avoids flooding and lessens transmission overhead. Shortening the scope-interval results more NRL in FSR-Mod than FSR-Orig.

%\vspace{-0.5cm}
\section{Conclusion and Future Work}
%In WMhNs, routing becomes a challenging issue when no. of nodes or network loads increase. Routing protocols in this regard, perform an efficient role to calculate routes.
In this paper, we have evaluated and compared the performance of three widely used proactive protocols; DSDV, FSR and OLSR. Total routing load attained by a protocol is based upon two factors; control traffic generated by control packets and data traffic forwarded through routes of non-optimal path lengths. Therefore, for evaluating the routing efficiency of these protocols in dense networks and with different data traffic rates, we have taken different scalabilities and varying data loads. For the analysis, three performance parameters; E2ED, NRL and throughput are computed by using NS-2. Finally, we observed that OLSR is more scalable because of reduction of routing overhead due to MPRs, as OLSR allows retransmission through MPRs. On the other hand, FSR is more suitable for high network loads due to scope routing through GF (no flooding), which reduces broadcasting storm, thus saves, more bandwidth and achieves high throughput when data traffic increase.

In future, we are interested to minimize energy consumed during routing by optimize these routing techniques both at MAC and network layer, like in, [11], [12], and [13].

\end{document}